\documentclass{ws-ijmpa}

\begin{document}

\markboth{T. Pietrycki}
{Direct photon production in $pp$ and $p \bar p$ collisions at high energies}

%
\catchline{}{}{}{}{}
%

\title{DIRECT PHOTON PRODUCTION IN $pp$ and $p \bar p$ 
COLLISIONS \\ AT HIGH ENERGIES
\footnote{Based on talk at MESON06 (Krak\'ow, 9 - 13 June 2006)}
}

\author{TOMASZ PIETRYCKI}

\address{Institute of Nuclear Physics\\
PL-31-342 Cracow, Poland\\
Tomasz.Pietrycki@ifj.edu.pl}

\author{ANTONI SZCZUREK}

\address{Institute of Nuclear Physics\\
PL-31-342 Cracow, Poland\\
and\\
University of Rzesz\'ow\\
PL-35-959 Rzesz\'ow, Poland\\
Antoni.Szczurek@ifj.edu.pl}

\maketitle

\begin{history}
\received{Day Month Year}
\revised{Day Month Year}
\end{history}

\begin{abstract}

The invariant cross sections for direct photon production in
hadron-hadron collisions are calculated for several initial energies
(SPS, ISR, S$p \bar p$S, RHIC, Tevatron, LHC) including initial parton
transverse momenta within the formalism of unintegrated
parton distributions (UPDF).
Kwieci\'nski UPDFs provide very good description of all world data,
especially at SPS and ISR energies.
Inclusion of the QCD evolution effects and especially
their effect on initial parton transverse momenta allowed to solve
the long-standing problem of understanding
the low energy and low transverse momentum data.

\end{abstract}

\ccode{PACS numbers: 12.38.Bx, 13.95.Qk, 13.60.Hb}

\section{Introduction}	

It was realized relatively early that the transverse momenta of initial
(before a hard process) partons may play an important role in order to
understand the distributions of produced direct photons, especially
at small transverse momenta (see e.g. Ref.~\refcite{Owens}). 

The simplest way to include parton transverse momenta is via
Gaussian smearing \cite{Owens}\cdash\cite{AM04}.
This phenomenological approach is not completely justified theoretically. 

The unintegrated parton distribution functions (UPDFs) are
the basic theoretical quantities that take into account explicitly 
the parton transverse momenta.
The UPDFs have been studied recently in the context of different
high-energy processes \cite{Gribov_Levin_Ryskin}\cdash\cite{LS05}.
These works concentrated
mainly on gluon degrees of fredom which play the dominant role
in many processes at very high energies. At somewhat lower energies
also quark and antiquark degrees of freedom become equally important.
Recently the approach which dynamically includes transverse momenta
of not only gluons but also of quarks and antiquarks was applied
to direct-photon production \cite{LZ05,KMR_photons}.

Up to now there is no complete agreement how to include evolution
effects into the building blocks of the high-energy processes --
the unintegrated parton distributions. In Ref.~\refcite{PS06}
we have discussed in detail a few approaches how
to include transverse momenta of the incoming partons in order
to calculate distributions of direct photons.

\section{Unintegrated parton distributions}

In general, there are no simple relations between unintegrated
and integrated parton distributions.
Some of UPDFs in the literature are obtained based on familiar
collinear distributions, some are obtained by solving evolution
equations, some are just modelled or some are even parametrized.
A brief review of unintegrated gluon distributions (UGDFs) 
can be found in Ref.~\refcite{LS06}.

In some of the approaches mentioned above one imposes the following
relation between the standard collinear distributions and UPDFs:
\begin{equation}
a(x,\mu^2) = \int_0^{\mu^2} f_a(x,k_t^2,\mu^2)
\frac{d k_t^2}{k_t^2}   \; ,
\end{equation}
where $a = xq$ or $a = xg$.

Since familiar collinear distributions satisfy sum rules, one can
define and test analogous sum rules for UPDFs.
We have discussed this issue in more detail in Ref.~\refcite{PS06}.
Some other approaches for UPDFs are discussed e.g. in Ref.~\refcite{LS06}.

\section{UPDFs and photon production}

\begin{figure}[!htb]
\includegraphics[width=.6\textwidth]{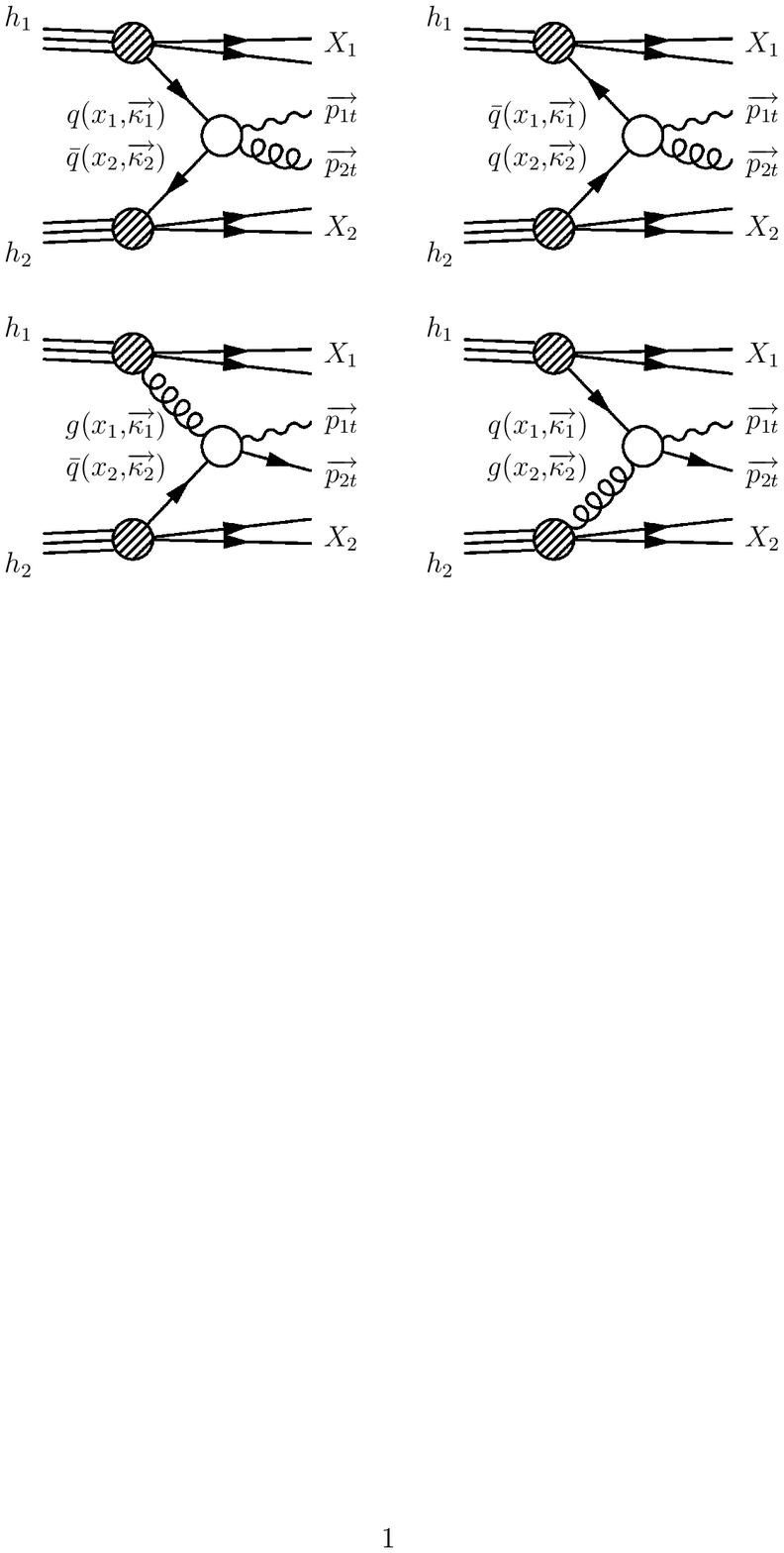}
\caption[*]{
The diagrams included in our $k_t$-factorization approach 
with the notation of kinematical variables.
\label{fig:diagrams}
}
\end{figure}

The cross section for the production of a photon and an associated parton (jet)
can be written as
\begin{eqnarray}
\frac{d\sigma(h_1 h_2 \rightarrow \gamma, parton)}
{d^2p_{1,t}d^2p_{2,t}} &=& \int dy_1 dy_2
\frac{d^2 k_{1,t}}{\pi}\frac{d^2 k_{2,t}}{\pi}
\frac{1}{16\pi^2(x_1x_2s)^2} \; \sum_{i,j,k}
\overline{|\mathcal{M}(i j \rightarrow \gamma k)|^2}
\nonumber \\
&\cdot&\delta^2(\vec{k}_{1,t}
+\vec{k}_{2,t}
-\vec{p}_{1,t}
-\vec{p}_{2,t})
f_i(x_1,k_{1,t}^2)f_j(x_2,k_{2,t}^2) \; ,
\label{basic_formula}
\end{eqnarray}
where
$\vec{k}_{1,t}$ and $\vec{k}_{2,t}$ are transverse
momenta of incoming partons. In the leading-order approximation: 
$(i,j,k) = (q,{\bar q},g), ({\bar q},q,g), (q, g, q), (g, q, q)$, 
etc. (see Fig.~\ref{fig:diagrams}).

\begin{figure}[htb]
\includegraphics[width=.6\textwidth]{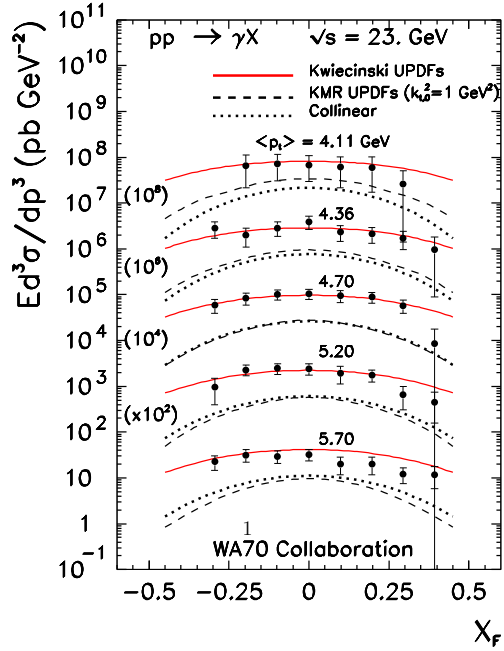}
\caption[*]{
Invariant cross section for direct photons for $\sqrt s$ = 23 GeV 
as a function of Feynman $x_F$ for different bins 
of transverse momenta. In this calculation off-shell 
matrix elements for subprocesses with gluons were used.
The Kwieci\'nski UPDFs were calculated with the factorization 
scale $\mu^2$ = 100 GeV$^2$. The theoretical results 
are compared with the WA70 collaboration data
\cite{data_WA70}.
\label{fig:WA70}
}
\end{figure}

If one makes the following replacements
\begin{equation}
f_i(x_1,k_{1,t}^2) \rightarrow x_1p_i(x_1)
\delta(k_{1,t}^2)
\end{equation}
and
\begin{equation}
f_j(x_2,k_{2,t}^2) \rightarrow x_2p_j(x_2)
\delta(k_{2,t}^2)
\end{equation}
then one recovers the standard collinear formula
(see e.g.\refcite{Owens}).

The inclusive invariant cross section for direct photon production can be
written
\begin{eqnarray}
\frac{d \sigma(h_1h_2 \to \gamma)}{dy_1 d^2p_{1,t}}
&=& \int dy_2 \frac{d^2 k_{1,t}}{\pi} \frac{d^2 k_{2,t}}{\pi}
\left( ... \right) |
_{\vec{p}_{2,t} = \vec{k}_{1,t} + \vec{k}_{2,t} -
  \vec{p}_{1,t}} \nonumber
\\
&=& \int d k_{1,t} d k_{2,t} \;
I(k_{1,t},k_{2,t};y_1, p_{1,t}) \; .
\label{inclusive1}
\end{eqnarray}
The integrand $I(k_{1,t},k_{2,t};y_1,p_{1,t})$ defined above depends 
strongly on the UPDFs used.

\begin{figure}[!htb] 
\begin{center}
\includegraphics[width=.45\textwidth]{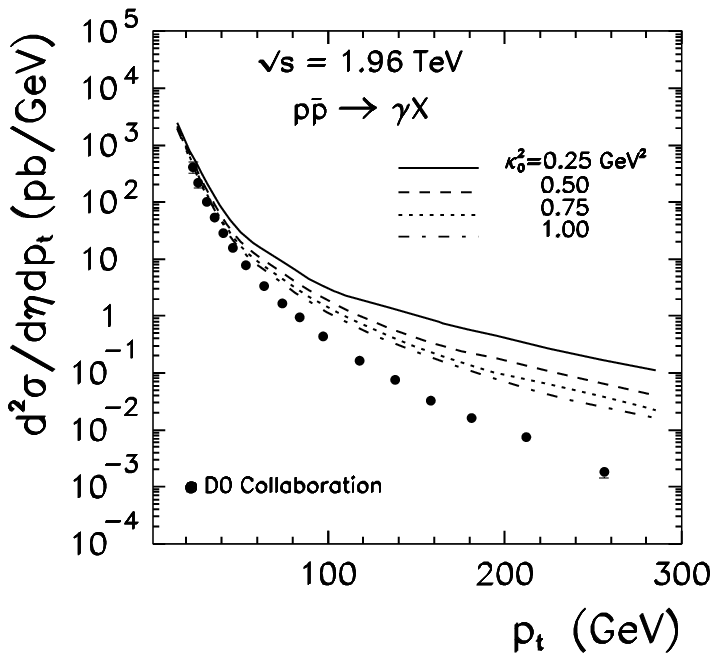}
\includegraphics[width=.45\textwidth]{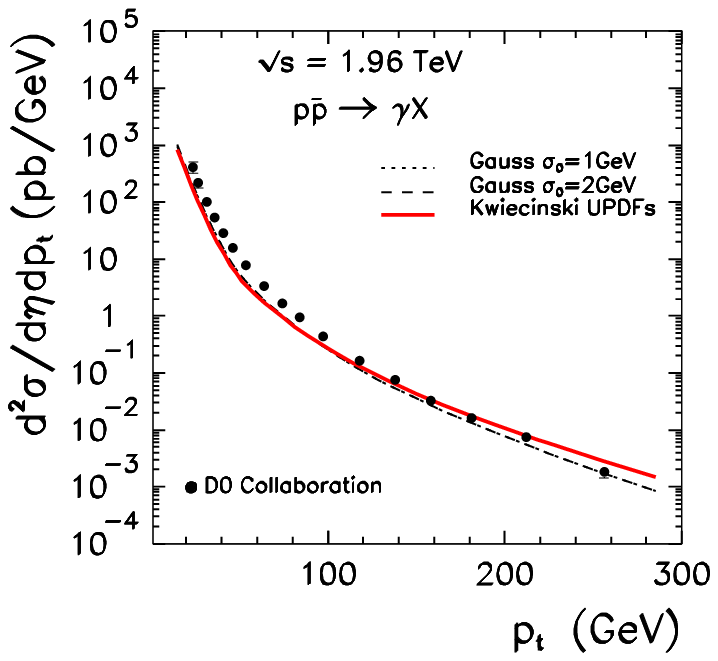}
\caption[*]{
Cross section for direct photons for
$\sqrt s = 1.96$ TeV. In this calculation off-shell matrix element for
gluons were used.
(a) standard KMR prescription,
(b) Gaussian smearing ($\sigma_0$ = 1,2 GeV) versus Kwieci\'nski
UPDFs.
The experimental data are from \cite{data_D0_w1960}. 
\label{fig:inv_cs_w1960}
}
\end{center}
\end{figure}

In Fig.~\ref{fig:inv_cs_w1960} we show 
inclusive invariant cross section
as a function of Feynman $x_F$ for several experimental values of
photon transverse momenta as measured by the WA70 collaboration.

It is well known that the collinear approach (dotted line) fails to
describe the low transverse momentum data by a sizeable factor of 4 or
even more. Also the $k_t$-factorization result with the KMR UPDFs
(dashed line) underestimate the low-energy data. In contrast,
the Kwieci\'nski UPDFs (solid line) describe the WA70 collaboration
data almost perfect \cite{data_WA70}.

In Fig.~\ref{fig:inv_cs_w1960} we compare results obtained with
different UGDFs with a recent experimental data of the D0 collaboration.
The unintegrated parton distribution approach with the KMR UPDFs is
clearly inconsistent with the standard collinear approach at large
transverse momenta. This is caused by the presence of large-$k_t$ tails
(of the 1/$k_t$ type) in the KMR UPDFs.
It is not the case for the Gaussian and Kwieci\'nski UPDFs which 
seem to converge to the standard collinear result at large photon
transverse momenta. In this respect the latter UPDFs seems preferable.


\end{document}